\DocumentMetadata{lang=en-GB,tagging=on}
\documentclass[a4paper]{article}
\usepackage[
unicode,
pdfauthor={Leron Borsten, 金炯錄},
pdftitle={Limits to Computational Acceleration Imposed by Quantum Field Theory and Quantum Gravity},
pdfsubject={hep-th,gr-qc},
breaklinks=true,
final
]{hyperref}
\usepackage[utf8]{inputenc}
\usepackage[T1,T2A]{fontenc}
\usepackage[unicode,final]{hyperref}
\usepackage{amsmath,amssymb,amsthm,orcidlink,cleveref,tikz-cd,mleftright}
\usepackage[noadjust,nosort]{cite}

\usetikzlibrary{babel,calc,arrows.meta,decorations.markings,decorations.pathmorphing}
\tikzset{
    horizon/.style={thick,dashed},
    singularity/.style={thick,
        decorate,
        decoration={snake, amplitude=0.2em, segment length=4mm}
    },
    region/.style={font=\small},
    arrow/.style = {
        postaction={decorate},
        decoration={
            markings,
            mark=at position 0.5 with {\arrow[scale=2]{latex}}
        }
    },
}
\usepackage{pgfplots}
\usepackage{CJKutf8}
\usepackage[final]{microtype}
\usepackage[latin,russian,german,italian,british]{babel}
\usepackage[osf]{newpxtext}
\usepackage[varbb,slantedGreek]{newpxmath}
\usepackage[artemisia]{textgreek}
\renewcommand\pi\piup

\begin{document}
\title{Limits to Computational Acceleration Imposed by Quantum Field Theory and Quantum Gravity}
\author{Leron Borsten\textsuperscript{\orcidlink{0000-0001-9008-7725}}\footnote{Centre for Mathematics and Theoretical Physics Research, Department of Physics, Astronomy and Mathematics, University of Hertfordshire, Hatfield, Hertfordshire\ \textsc{al10 9ab}, United Kingdom}\\Hyungrok Kim~(\begin{CJK*}{UTF8}{bsmi}金炯錄\end{CJK*})\textsuperscript{\orcidlink{0000-0001-7909-4510}}\footnotemark[1]
\\ [1em]
\tt
\{\href{mailto:l.borsten@herts.ac.uk}{l.borsten},\,\href{mailto:h.kim2@herts.ac.uk}{\texttt{h.kim2}}\}@herts.ac.uk
}
\maketitle
\begin{abstract}
A computer, in order to perform a given computation, requires a certain amount of space (memory) and a certain amount of time (runtime). This leaves certain computations beyond reach due to technological limits on processing speed and  memory density. Some  computations, such as the halting problem, are not possible even in principle. However, curved spacetimes and exotic fields appear to provide avenues to accelerate computation, for instance by exploiting time dilation. Impossible computations seemingly become tractable, butting up against intuition. However, 
we show that such schemes are consistently thwarted by physical effects from quantum gravity (including swampland conjectures) and quantum field theory in curved space.
More precisely, we show that an observer and a computer able to withstand energy scales up to order \(E\) can, by using relativistic effects, accelerate computation at a rate of at most \(\mathcal O(1)E\) e-folds per unit time in natural units: \((\ln\alpha)/\tau\lesssim E\).
The Bekenstein bound for entropy can then be understood as the space (memory) analogue to (run)time: if a computer of length scale \(D\), operating at energies up to order \(E\), has access to \(N\) different memory states, then \((\ln N)/D\lesssim E\).
\end{abstract}
\tableofcontents

\section{Introduction and Summary}
\paragraph{Background.}
It is an old idea that fundamental physics has close links to the theory of computation: information is physical. For instance, recently it has been suggested \cite{Brown:2015bva} that quantum computational complexity is related to the action of a certain patch of bulk spacetime in the AdS/CFT correspondence, and computational complexity theory has been used to argue against the firewall proposal \cite{Harlow:2013tf}. There are by now various connections between computation and AdS/CFT  \cite{Almheiri:2014lwa,Freedman:2016zud,Aaronson:2022iyo}. Even prior to this, many models of computation are implicitly or explicitly based on what physics allows. Indeed, the original definition of a Turing machine was motivated by considering the kinds of computation that a physical machine realisable in classical mechanics can accomplish; similarly, quantum circuits and related models of quantum computation are motivated by idealised machines that can be realised in quantum mechanics. For more on physically motivated models of computation, see \cite{Perales-Eceiza:2024qhd} and references therein.

In most models of computation, in order to compute something, the computer needs access to space (in the form of memory space) as well as time (that is, the runtime duration during which a computation is performed); the amount of space and time a given computation requires can be quantified, leading to complexity classes such as \(\mathsf{PSPACE}\) (computational problems that, for a classical Turing machine, require an amount of memory that grows at most polynomially in relation to the size of the input) and \(\mathsf{P}\) (computational problems that, for a classical Turing machine, require an amount of time that grows at most polynomially in relation to the size of the input). In one extreme, if the amount of memory given to a computer is finite, even if it has access to arbitrary amounts of runtime, that computer is \emph{not} capable of universal computation and reduces to a finite automaton with strictly weaker computational power. On another extreme, consider a hypothetical computer (`hypercomputer') which has access to an actually infinite (rather than arbitrarily long) runtime: the first computational step takes time \(\frac12t_0\), the second step takes time \(\frac14t_0\), the third \(\frac18t_0\), and so on, so that an infinite number of computational steps can take place in a finite time \(t_0\) \cite{ord2002hypercomputationcomputingturingmachine}. Such a computer can compute things that an ordinary computer cannot: for instance, while ordinary computers cannot solve the halting problem of determining whether a given program terminates, such a hypercomputer can  determine whether a given program terminates simply by running it.

While such idealisations are seemingly implausible,  a natural question is whether one can in fact improve computational power  by means of relativistic and other physical effects. For instance, one can try to exploit time dilation to increase the runtime available to a computer, or use exotic forms of memory possible with suitable field content. In particular, it has been claimed \cite{pitowsky,1992FoPhL...5..173H,Earman_Norton_1993,Hogarth_1994,earman1996infinite,Etesi:2001ze,Manchak2010} (see the survey \cite{sep-spacetime-supertasks}) that classical general relativity allows \emph{infinite} compression of time, making an infinite amount of runtime available to a computer and thus realising hypercomputation. For example, solutions to Einstein's equations known as Malament--Hogarth spacetimes admit infinite proper time  geodesics that intersect finite proper time geodesics. To perform a hypercomputation, just send your everyday computer along the infinite proper time  time geodesic and read-off the result when you meet it  again.  Malament--Hogarth spacetimes include rather pedestrian examples:   anti-de~Sitter space and Kerr--Newmann black holes. This makes it hard to accept that they should be ruled-out by some kind of computational curtailment conjecture. 

Of course, one can raise many objections as to the physical realisability of such hypercomputers \cite{Earman_Norton_1993,earman1996infinite},  avoiding the need to restrict  by fiat. Here, we consider fundamental obstructions to constructing hypercomputers using spacetime geometry and exotic fields. In all cases, quantum effects limit what is possible and point towards a fundamental bound on computational acceleration. 

\paragraph{Results.}
In this paper, we observe that effects coming from quantum field theory on curved space and quantum gravity (including swampland conjectures) limit the extent to which one can increase runtime and memory in this way, in particular ruling out hypercomputation by means of Malament--Hogarth spacetimes. In detail, we obtain the following results.

\medskip
\noindent\emph{\textbf{Compressing time.}}
Suppose that, during a proper time \(\tau_\mathrm{obs}\) of the observer, the computer can access a runtime of \(\tau_\mathrm{comp}\) due to relativistic effects. We define the \emph{time advantage} as the ratio
\begin{equation}
    \alpha = \frac{\tau_\mathrm{comp}}{\tau_\mathrm{obs}}.
\end{equation}
That is, if \(\alpha\) is large, we have succeeded in speeding up the computer by a factor of \(\alpha\). In \cref{sec:time}, we show that, in Minkowski space, (anti-)de Sitter space and many other physically realistic spacetimes, if the observer and the computer can survive energies up to \(E\), then one can only obtain \(\mathcal O(1)E\) many e-folds of time advantage per unit proper time of the observer, i.e.
\begin{equation}\label{eq:unruh-bound-intro}
    \frac{\ln\alpha}{\tau_\mathrm{obs}} \le\mathcal O(1)E,
\end{equation}
in the limit \(\tau_\mathrm{obs}\to\infty\). In particular, if the energy scale \(E\) is specified as a temperature \(T\) beyond which the observer is destroyed, then in Minkowski space the \(\mathcal O(1)\) constant is found to be
\begin{equation}
    \frac{\ln\alpha}{\tau_\mathrm{obs}} \le \pi T.
\end{equation}
Since in Minkowski space the bound arises from the Unruh effect, we call this the \emph{Unruh bound}.
However, surprisingly, similar bounds hold in a variety of physically realistic spacetimes and due to a variety of relativistic quantum effects: the Unruh effect, Hawking radiation, the Casimir effect, and the no-transmission principle of Engelhardt and Horowitz \cite{Engelhardt:2015gla}.

\medskip
\noindent\emph{\textbf{Compressing space.}}
Suppose a computer has size on the order of \(D\) and has energy \(E\). The Bekenstein bound \cite{Bekenstein:1980jp} implies that the number \(N\) of possible memory states of the computer is bounded above by
\begin{equation}
    \frac{\ln N}D \le \mathcal O(1)E.
\end{equation}
In particular, the bound should be saturated if the memory is in the form of a Schwarzschild black hole. If \(D\) is the diameter of the Schwarzschild black hole and \(E\) its mass, then for all dimensions \(d\) where Schwarzschild black holes exist in flat space, we have
\begin{equation}\label{eq:bekenstein-bound-intro}
    \frac{\ln N}D \le \pi E.
\end{equation}
This bound \eqref{eq:bekenstein-bound-intro} bears a striking resemblance to the bound \eqref{eq:unruh-bound-intro}.\footnote{See \cite{Hayden:2023ocd} for another discussion of the relation between the Unruh effect and the Bekenstein bound.}
This suggests that \(N\) should be seen as an analogue of the time advantage \(\alpha\), just as the diameter \(D\) is an analogue of the proper time \(\tau_\mathrm{obs}\).

Various swampland conjectures should hold in order for the Bekenstein bound to be satisfied. In particular, the species-scale conjecture \cite{Palti:2019pca} states that one cannot exceed the Bekenstein bound \eqref{eq:bekenstein-bound-intro} by means of an unbounded number of particle speeds to encode information in a finite spatial region. Similarly, the (refined) distance conjecture \cite{Ooguri:2006in,Blumenhagen:2018nts}, together with a plausible bound \eqref{eq:precision} on the precision with which one can measure moduli, implies that one cannot exceed \eqref{eq:bekenstein-bound-intro} by means of storing information in moduli.

\paragraph{Limitations.}
In our discussion, we assume that both the observer and the computer are macroscopic systems well described by effective field theories below some energy scale; in particular, they are always `embodied'.
One can circumvent some of our conclusions if one considers an `observer' purely made of information.
For instance, suppose that the observer's mind is made of purely classical information, so that a brain-scanning machine can
extract the classical information (and perhaps destroy the body). This classical information can then be transmitted at the speed of light, so that the `observer' effectively follows a timelike trajectory, `experiencing' zero elapsed proper time.
Even if the observer is an intrinsically quantum system, by means of enough entanglement, one can imagine quantum-teleporting the observer, once again at the speed of light (see \cite{Horodecki:2009zz} for a review).
The scope of the present paper excludes these scenarios.

\paragraph{Organisation.} This paper is organised as follows. We briefly review requirements for space and time in computability and computational complexity theory in \cref{sec:review}. Then we discuss what we call the Unruh bound \eqref{eq:unruh-bound-intro} for how much the runtime for a computer may be compressed in \cref{sec:time}, examining various cases in Minkowski space, (anti-)de Sitter space, and black holes. \Cref{sec:space} discusses how the Bekenstein bound forms a parallel to the Unruh bound, describing how much one can compress space (memory) for a computer in a parallel form.

\paragraph{Conventions.} We work in \(d\) spacetime dimensions with a mostly-plus metric. We work in units where \(c=\hbar=1\), but we do not set the gravitational constant to unity. The latter is rather unnatural from the point of view of string theory: the \(d\)-dimensional gravitational constant is not fundamental but is determined by the moduli of a given string compactification, such as the volume of the internal dimensions and the vacuum expectation value of the dilaton.
The Planck energy \(M_\mathrm{Pl}\) in \(d\) dimensions is unrationalised, so that \(M_\mathrm{Pl} = G_\mathrm N^{-1/(d-2)}\) where \(G_\mathrm N\) is Newton's gravitational constant.

\section{Informal review of computation and supertasks}\label{sec:review}
A \emph{Turing machine} is a simple model of computation.
Its state space is \(\Gamma^{\mathbb Z}\times Q\times\mathbb Z\), where \(\Gamma^{\mathbb Z}\) is the state of an infinite-one-dimensional tape with cells \(x_i\) labelled by integers \(i\) with \(\Gamma\) a finite set of possible states of the cell, plus an internal state \(q\in Q\), where \(Q\) is a finite set of possible internal states, and \(\mathbb Z\) labels the position of the head on the tape. Its evolution is deterministic, given a transition function that, based on the value \(x_i\) stored on the current cell and the current internal state \(q\in Q\), determines whether the head moves forward or backward, how the internal state is updated, and whether to halt. It is a remarkable fact that there exist \emph{universal Turing machines}, i.e.\ Turing machines that can simulate any other Turing machine, and that universal Turing machines are equivalent in computational power to many other natural models of computation, such as the lambda calculus or the Post system, in the sense that a function \(\mathbb N\to\{0,1\}\) (or, equivalently, a real number\footnote{
    There exist natural bijections between the function space \(\{0,1\}^{\mathbb N}\) and \(\mathbb R\).
    For example: \(\mathbb R\) can be mapped via \(x\mapsto\pi^{-1}\tan(x)+1/2\) to the open interval \((0,1)\). An element \(x\in(0,1)\) determines its sequence of binary digits \(x=2^{-1}f(1)+2^{-2}f(2)+\dotsb+2^{-n}f(n)\), which can be regarded as a function \(f\colon\mathbb N\to\{0,1\}\).
}) is computable by a universal Turing machine if and only if it is computable by the lambda calculus, etc. (Note that we are concerned with \emph{computability}, i.e.\ whether a function can be computed at all, rather than \emph{computational complexity}, i.e.\ how long or how much space does it take to approximate a computable real number to a desired degree of precision.)

It is another remarkable fact that these natural models of computation are equivalent in computability strength to many models that naturally embed into physics. For instance, a model of computation based on classical billiards is equivalent in computability strength to universal Turing machines \cite{miranda2025classicalbilliardscompute}. Similarly, quantum circuits, a model of computation based on quantum computers, are also equivalent in computability strength to universal Turing machines (although they are suspected to be \emph{not} equivalent in terms of computational complexity, i.e.\ the class of problems solvable within `reasonable time' by quantum circuits, namely \(\mathsf{BQP}\), is conjectured to be strictly larger than the corresponding class for Turing machines, namely \(\mathsf{P}\)). This raises the question of what \cite{cotogno,piccinini} calls the physical Church--Turing thesis and \cite{Etesi:2001ze} calls the Church--Kalmár--Turing thesis, namely the assertion that the class of real numbers computable by (systems embedded in) a given physical theory \(\mathcal T\) is equivalent to those computable by Turing machines. There is a different version of the physical Church--Turing thesis for each physical theory \(\mathcal T\), which can have different truth values.

Note that the computational complexity version of the Church--Turing thesis, namely that the class of functions efficiently computable by the physical theory \(\mathcal T\) is equivalent to \(\mathsf{P}\), is widely conjectured to be \emph{false} \cite{Ethan:2006gnj}: namely, quantum mechanics can efficiently compute functions or real numbers that Turing machines cannot.

\section{Limits on compressing time}\label{sec:time}
Suppose that you want to compute a task that normally takes time \(T\) to compute on a given computer. You have a brilliant idea: you travel along a worldline \(\gamma_\mathrm{obs}\), while the computer travels along a worldline \(\gamma_\mathrm{comp}\) with the same endpoints. In a curved spacetime, one can arrange for the proper time \(\tau_\mathrm{obs}\) along \(\gamma_\mathrm{obs}\) to be much shorter than the proper time \(\tau_\mathrm{comp}\) along \(\gamma_\mathrm{comp}\); this seems to have accelerated the computation by a factor of
\begin{equation}
    \alpha=\frac{\tau_\mathrm{comp}}{\tau_\mathrm{obs}}=\frac{\int_{\gamma_\mathrm{comp}}\mathrm d\tau}{\int_{\gamma_\mathrm{obs}}\mathrm d\tau},
\end{equation}
where \(\mathrm d\tau\) is the infinitesimal element of proper time; we call this factor \(\alpha\) the \emph{time advantage}.

\begin{figure}
\begin{center}
\begin{tikzpicture}[scale=1.2]

  \coordinate (x) at (0,0);
  \coordinate (y) at (1.2,2);
  \coordinate (z) at (1.2,2.5);

  \draw[thick]
    (x) .. controls (0.3,1) and (0.7,1.6) .. (y)
    node[midway, right] {$\gamma$};

  \draw[dashed]
    (z) -- ++(-2,-2);
  \draw[dashed]
    (z) -- ++(2,-2);

  \fill (x) circle (2pt);
  \fill (y) circle (2pt);
  \fill (z) circle (2pt);

  \node[below left] at (x) {$x$};
  \node[left] at (y) {$y$};
  \node[above right] at (z) {$z$};

\end{tikzpicture}
\end{center}
\caption{In a Malament--Hogarth spacetime, a computer travels along a timelike curve \(\gamma\) whose proper length is infinite. Its endpoint \(y\) is in the causal past of the observer located at \(z\), who can then observe the results of a computation that takes infinitely long to compute, e.g.\ whether a Turing machine ever halts.}\label{fig:malament-hogarth-spacetime}
\end{figure}
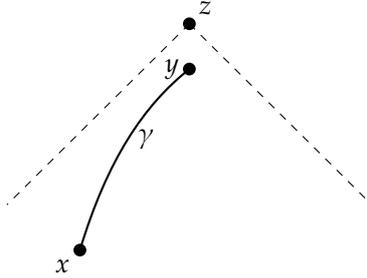

In particular, consider the case where \(\int_{\gamma_\mathrm{comp}}\mathrm d\tau=\infty\) while \(\int_{\gamma_\mathrm{obs}}\mathrm d\tau<\infty\), so that one has an infinite time acceleration factor. A spacetime with such a worldline is called a Malament--Hogarth spacetime \cite{1992FoPhL...5..173H,Earman_Norton_1993} and provides a putative counterexample to the physical Church--Turing thesis for classical general relativity. A Malament--Hogarth spacetime is simply a pseudo-Riemannian manifold \((M,g)\) with a timelike half-curve \(\gamma\colon[0,\infty)\to M\) with infinite proper length such that (the image of) \(\gamma\) lies in the causal past of some point \(x\in M\). Suppose, for instance, that one wants to determine whether a Turing machine \(T\) halts. One simply takes a computer that moves along \(\gamma\) and simulates \(T\); if \(T\) is found to halt at some finite time \(t\in[0,\infty)\), then the computer sends out a signal. An observer at \(x\) can then tell, with certainty, whether \(T\) halts: if they observe a signal, then \(T\) halts; if they do not observe a signal, then \(T\) does not halt (\cref{fig:malament-hogarth-spacetime}).
A Malament--Hogarth spacetime necessarily fails to be globally hyperbolic \cite{1992FoPhL...5..173H,Earman_Norton_1993}. On the other hand, many quite familiar spacetimes are Malament--Hogarth: for instance, anti-de~Sitter space \(\operatorname{AdS}_d\) and non-extremal Kerr--Newmann black holes are Malament--Hogarth \cite{Etesi:2001ze}.

Thus, if an asymptotically  \(\operatorname{AdS}_d\) spacetime were physically possible, it would naïvely seem to provide a means to solve the halting problem and disprove the physical Church--Turing thesis. This is unsettling to say the least; anti-de Sitter space is a rather benign solution of the Einstein equation with negative cosmological constant and under the AdS/CFT correspondence is dual to certain conformal field theories on its boundary, which are as benign as field theories can get.
This reasoning is, of course, too quick. We have exploited one pillar  of fundamental physics,  general relativity, but forgotten to account for the other, quantum field theory.   Below, we examine how quantum effects frustrate attempts to exploit Malament--Hogarth spacetime and, in particular, impose the bound \eqref{eq:unruh-bound-intro} on time advantage.

\subsection{Compressing runtime in Minkowski space}
\begin{figure}
\begin{center}
\begin{tikzpicture}[scale=2]
  \coordinate (x) at (0,0);
  \coordinate (y) at (1.75,3.5);
  \coordinate (z) at (0,3.5);
  \coordinate (y') at (1,2);
  \coordinate (z') at (0,3);
  \draw[thick,  arrow] (x) -- (z) node[midway, left=1ex] {\shortstack{observer's\\worldline \(\gamma_\mathrm{obs}\)}};
  \draw[thick,  arrow] (x) -- (y) node[midway, right=1ex] {\shortstack{computer's\\worldline \(\gamma_\mathrm{comp}\)}};
  \draw[dashed, arrow] (y') -- (z') node[near end, right=1ex] {photon};

  \fill (x) circle (1pt);

\end{tikzpicture}
\end{center}
\caption{In Minkowski space, due to time dilation, an observer observes a computer moving relative to them as running slower, not faster.}\label{fig:time-dilation}
\end{figure}
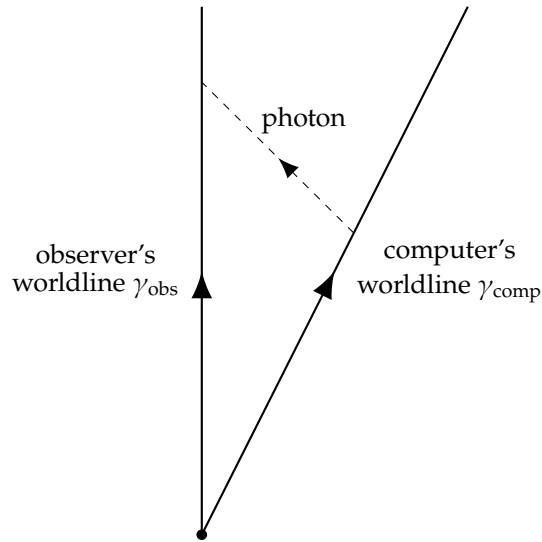

In Minkowski space, one cannot experience any speedup if both the observer and the computer move along geodesics due to the fact that special relativity always produces time \emph{dilation} (and length contraction): if both the observer as well as the computer follow timelike geodesics, communicating by lightlike or timelike signals, any relative velocity between the observer and the computer makes the computer \emph{less efficient} for the observer, not more efficient (\cref{fig:time-dilation}).

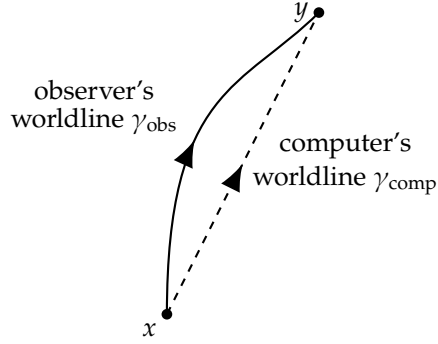
\begin{figure}
\begin{center}
\begin{tikzpicture}[scale=2]
  \coordinate (x) at (0,0);
  \coordinate (y) at (1,2);

  \draw[thick, arrow] (x) .. controls (0,1.5) and (0.5,1.5) .. (y) node[midway, left=1em] {\shortstack{observer's\\worldline \(\gamma_\mathrm{obs}\)}};
  \draw[thick, dashed, arrow] (x) -- (y) node[midway, right] {\shortstack{computer's\\worldline \(\gamma_\mathrm{comp}\)}};

  \fill (x) circle (1pt);
  \fill (y) circle (1pt);

  \node[below left] at (x) {$x$};
  \node[left] at (y) {$y$};

\end{tikzpicture}
\end{center}
\caption{In order to accelerate computation in Minkowski space, the observer undergoes uniform acceleration along a timelike curve \(\gamma_\mathrm{obs}\) while the computer follows a geodesic \(\gamma_\mathrm{comp}\) between the same endpoints.}\label{fig:minkowski}
\end{figure}

Of course, if the observer is allowed to deviate from geodesics by accelerating, time compression can occur --- an instance of the famous twin paradox.
In order to maximise the time advantage, the optimal situation is for the observer to undergo uniform acceleration at \(a=2\pi\Lambda\) while the computer moves along a geodesic, so that the velocity changes as much as possible, as shown in \cref{fig:minkowski}. (Note that the \emph{observer} must accelerate, not the computer: in the twin paradox, the accelerating twin ages less. Furthermore, we don't have a lightlike signal propagation from the computer to the observer, unlike \cref{fig:time-dilation}: this would amount to an infinite acceleration of the computer, but acceleration for the computer is detrimental. The assumption that the observer has finite mass is crucial; otherwise one could consider an `observer' who moves along a lightlike geodesic and be reflected off of a mirror, experiencing zero proper time.)
We may, without loss of generality, work in the frame where initially the observer is at rest. Then the observer's uniformly accelerated motion is
\begin{equation}
    (t,x) = (a^{-1}\sinh(a\tau),a^{-1}\cosh(a\tau)),
\end{equation}
where we ignore all but only one spatial coordinate in addition to the time coordinate. The observer starts at \((0,a^{-1})\) and, after a proper time \(\tau_\mathrm{obs}\), arrives at \(a^{-1}\sinh(a\tau_\mathrm{obs}),a^{-1}\cosh(a\tau_\mathrm{obs})\); the computer starts and ends at the same endpoints but moves along a geodesic. Therefore the proper time \(\tau_\mathrm{comp}\) available to the computer is
\begin{equation}
\begin{aligned}
    \tau_\mathrm{comp}&=\int_{\gamma_\mathrm{comp}}\mathrm d\tau=a^{-1}\sqrt{\sinh^2(a\tau_\mathrm{obs})-(\cosh(a\tau_\mathrm{obs})-1)^2}\\
    &=\sqrt2a^{-1}\sqrt{\cosh(a\tau_\mathrm{obs})-1}.
\end{aligned}
\end{equation}
This yields a time advantage given by 
\begin{equation}
 \alpha=\frac{\tau_\mathrm{comp}}{\tau_\mathrm{obs}}=
    \frac{\sqrt{2(\cosh(2\pi T\tau_\mathrm{obs})-1)}}{a\tau_\mathrm{obs}}.
\end{equation}

However, when including quantum effects, the acceleration induces Unruh radiation with a temperature given by $a/2\pi$. Thus, the maximum time advantage for an observer that can withstand temperatures up to \(T=a/2\pi\) is
\begin{equation}
    \alpha=\frac{\tau_\mathrm{comp}}{\tau_\mathrm{obs}}=
    \frac{\sqrt{2(\cosh(2\pi T\tau_\mathrm{obs})-1)}}{2\pi T\tau_\mathrm{obs}}
    =
    1+\frac{(2\pi T\tau_\mathrm{obs})^2}{24}+\frac{(2\pi T\tau_\mathrm{obs})^4}{1920}
    +\dotsb.
\end{equation}
At large \(T\), this scales as
\begin{equation}
    \ln\alpha \sim \pi T\tau_\mathrm{obs}
    -\ln(\sqrt2\pi T\tau_\mathrm{obs})
    +\dotsb.
\end{equation}
In particular, we always have
\begin{equation}\label{eq:unruh-bound}
    \ln \alpha\le \pi T\tau_\mathrm{obs}.
\end{equation}
That is, an observer that can withstand temperatures up to \(T\) can accelerate computations that take time \(\tau_\mathrm{obs}\) by at most \(\pi T\tau_\mathrm{obs}\) e-folds.

\paragraph{Rotational motion}
Other kinds of accelerated worldlines are less efficient (for sufficiently long proper time). For instance, consider the case where the computer remains stationary while the observer moves along a circle at an angular velocity \(\omega\) as measured by the inertial computer, so that the observer traces out a trajectory
\begin{equation}
    (t,x,y) = (t,R\cos(\omega t),R\sin(\omega t))
\end{equation}
(with other spatial coordinates suppressed).
The observer experiences a proper acceleration of magnitude
\begin{equation}
    a = \frac{R\omega^2}{1-R^2\omega^2},
\end{equation}
so that an observer who can withstand temperatures up to \(T\) can have acceleration at most \(a\lesssim T\).\footnote{An observer in uniform circular motion experiences a bath of particles following a non-thermal spectrum \cite{Unruh:1998gq,Biermann:2020bjh}; in the ultrarelativistic limit \(v\lesssim c\), depending on the energy scale of the response function, the non-thermal spectrum has characteristic `temperature' \(T\) between 
\(a/(4\sqrt3)\le T\le a/(2\sqrt3)\).}
The proper time experienced by the accelerating observer is
\begin{equation}
    \mathrm d\tau_{\text{obs}} = \sqrt{1-R^2\omega^2}\,\mathrm dt
    =\frac{\mathrm dt}{\sqrt{1+aR}},
\end{equation}
so that
we obtain a time compression rate of
\begin{equation}
    \alpha=\frac{\mathrm dt}{\mathrm d\tau_{\text{obs}}}=\sqrt{1+aR},
\end{equation}
or
\begin{equation}  
    \ln\alpha = \frac12\ln(1+aR),
\end{equation}
which is \emph{not} proportional to the proper time \(\tau_\mathrm{obs}\) experienced by the observer, unlike \eqref{eq:unruh-bound}, and hence less efficient than a uniform acceleration for large \(\tau\).

More generally, the Casimir effect should occur for all compact spacetimes and rules out time-advantage schemes using compact spatial topology. We will meet an example of this in \cref{ssec:topo}.

\paragraph{Geodesics in de~Sitter space}
The bound \eqref{eq:unruh-bound} continues to hold for all spacetimes where Unruh-type temperatures can be computed by isometrically embedding into a higher-dimensional Minkowski space \(\mathbb R^{1,k}\) for sufficiently large \(k\), such as de~Sitter space\footnote{But not anti-de~Sitter space, which requires a $(2,k)$ signature embedding.} and the Schwarzchild black hole \cite{Deser:1997ri}.\footnote{But not every isometric embedding correctly computes the temperature; see \cite{Paston:2014sta} for counterexamples.}
For example, consider de~Sitter space \(\operatorname{dS}_d\), which embeds into \(\mathbb R^{1,d}=\{(z^0,\dotsc,z^d)\mid z^M\in\mathbb R\}\) as
\begin{equation}
    \operatorname{dS}_d = \left\{z^M\in\mathbb R^{1,d}\mid z^Mz_M = L^2\right\},
\end{equation}
where \(L\) is the de~Sitter radius. A uniformly accelerating observer in de~Sitter space traces out the trajectory of an accelerating observer in \(\mathbb R^{1,d}\) with constant proper acceleration \(a\ge1/L\) and, therefore, observes Unruh radiation with temperature \(T=a/2\pi\). In particular, a geodesic in de~Sitter space corresponds to \(a=1/L\), leading to the temperature \(T=(2\pi L)^{-1}\) of the de~Sitter horizon.

Concretely, the trajectory of an accelerating observer with constant acceleration \(a_{\operatorname{dS}} = \sqrt{a^2-L^{-2}}\) in \(\operatorname{dS}_d\) can be parameterised in embedding space coordinates \((z^0,\dotsc,z^d)\) as
\begin{equation}
\begin{split}
    z^0 &= a^{-1}\sinh(at),\\
    z^1 &= a^{-1}\cosh(at),\\
    z^2 &= \sqrt{L^2-1/a^2}, \\
    z^3=\dotsb=z^d &= 0,
\end{split}\end{equation}
with \(a\ge1/L\). A thermal bath of particles of temperature 
\begin{equation}
T=a/2\pi= \frac{1}{2\pi}\sqrt{a_{\operatorname{dS}}^2 + \frac{1}{L^2}}
\end{equation} is observed. We may, without loss of generality, consider this trajectory extending for \(-t_\mathrm{obs}\le t\le t_\mathrm{obs}\) for a proper length of \(2t_\mathrm{obs}\).

The computer instead travels along a geodesic trajectory between the same endpoints that maximises proper time. An arbitrary geodesic can be written without loss of generality as
\begin{equation}
\begin{split}
    z^0 &= L\sinh\frac tL,\\
    z^1 &= L\cosh\frac tL\cos\theta,\\
    z^2 &= L\cosh\frac tL\sin\theta,\\
    z^3=\dotsb=z^d &= 0,
\end{split}\end{equation}
where \(-t_\mathrm{comp}\le t\le t_\mathrm{comp}\). In order for the computer's geodesic to have the same endpoints as the observer's trajectory, we must have
\begin{equation}
\begin{aligned}
    L\sinh\frac{t_\mathrm{comp}}L&= a^{-1}\sinh(at_\mathrm{obs}),\\
    L\cosh\frac{t_\mathrm{comp}}L\cos\theta&=a^{-1}\cosh(at_\mathrm{obs}),\\
    L\cosh\frac{t_\mathrm{comp}}L\sin\theta&=\sqrt{L^2-a^{-2}}.
\end{aligned}
\end{equation}
In particular,
\begin{equation}
    t_\mathrm{comp}=L\sinh^{-1}\frac{\sinh(at_\mathrm{obs})}{aL}.
\end{equation}
Thus, the time advantage for an observer who can survive temperatures up to \(T=a/2\pi\) is
\begin{equation}
    \alpha = \frac{t_\mathrm{comp}}{t_\mathrm{obs}}
    = \frac{L\sinh^{-1}\mleft((2\pi TL)^{-1}\sinh(2\pi Tt_\mathrm{obs})\mright)}{t_\mathrm{obs}}.
\end{equation}
Suppose that we go to the limit where \(t_\mathrm{obs}\) is large, with \(a\) and \(L\) fixed. Then hyperbolic sine may be approximated as \(\sinh x\approx \frac12\exp x\), such that
\begin{equation}
    \alpha = \frac{L\sinh^{-1}\mleft((2\pi TL)^{-1}\sinh(2\pi Tt_\mathrm{obs})\mright)}{t_\mathrm{obs}} \approx 2\pi TL-(L/t_\mathrm{obs})\ln(2\pi TL).
\end{equation}
So, on time scales much longer than the de~Sitter radius, the time advantage \(\alpha\) approaches the constant \(2\pi TL\),
independent of the observer's time \(t\), which is much smaller than the bound \eqref{eq:unruh-bound}.

\subsection{Nontrivial topology  for accelerating computation}\label{ssec:topo}
One putative counterexample to \eqref{eq:unruh-bound} is the case of spacetime with compact spatial directions. For simplicity, we consider a two-dimensional spacetime \(\mathbb R\times\mathbb S^1=\{(t,x)|t\in\mathbb R,x\in\mathbb R/(2\pi R)\}\), where the first factor \(\mathbb R\) is (noncompact) time and the second factor \(\mathbb S^1\) corresponds to a periodic one-dimensional space with radius \(R\). The isometry group \(\operatorname{Isom}(\mathbb R\times\mathbb S^1)\) does \emph{not} map every timelike geodesic to every other geodesic, and there is a preferred notion of an inertial observer that is at rest with respect to the spacetime.

\begin{figure}\centering
\begin{tikzpicture}[scale=1]
  \draw[domain=0:900,smooth,variable=\t,samples=100, thick, arrow] plot ({2*sin(\t)},{cos(\t+180)+0.004*\t-1}) node [pos=1, right=5em] {\shortstack{observer's\\worldline \(\gamma_{\mathrm{obs}}\)}};

  \coordinate (x) at (0,-2);
  \coordinate (y) at (0,3.6);

  \draw[thick, dashed, arrow] (x) -- (y) node[pos=0.9, left] {\shortstack{computer's\\worldline \(\gamma_\mathrm{comp}\)}};
  \fill (x) circle (1pt);
  \fill (y) circle (1pt);
\end{tikzpicture}
\caption{In a cylindrical spacetime, one can obtain unboundedly high time advantage without any Unruh radiation, but the observer instead encounters relativistic Casimir energy--momentum.}\label{fig:cylinder}
\end{figure}
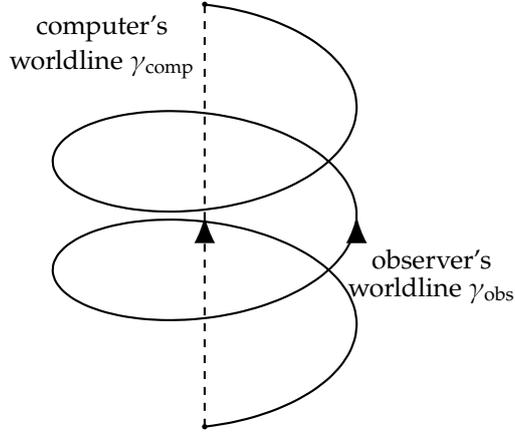

In this case, consider a setup (\cref{fig:cylinder}) where the computer remains at rest while the observer moves along space with an angular velocity \(\omega<1/R\) (as observed by an observer at rest):
\begin{align}
    \gamma_\mathrm{comp}(\tau)&=(\tau,0),&
    \gamma_\mathrm{obs}(\tau)&=\left(\tau/\sqrt{1-\omega^2R^2},\omega R\tau/\sqrt{1-\omega^2R^2}\right).
\end{align}
Both the computer and the observer travel from \((0,0)\) to \((2\pi N/\omega,0)\), where \(N\) is an integer. In this case, the time advantage coefficient is
\begin{equation}
    \alpha = \frac1{\sqrt{1-\omega^2R^2}},
\end{equation}
which is unbounded above as \(\omega\to1/R\). Unlike the previous cases, however, this is \emph{not} accompanied by any Unruh radiation: there is no curvature nor any acceleration, and an explicit computation in quantum field theory in curved space using Bogoliubov transformations makes it clear that there is no mixing between positive-energy and negative-energy modes and hence no Unruh radiation.\footnote{
    The cylindrical spacetime \(\mathbb R\times\mathbb S^1\) can be isometrically embedded into Minkowski space \(\mathbb R^{1,2}\) in an obvious way, and this maps \(\gamma_\mathrm{obs}\) to an observer undergoing uniform circular motion in Minkowski space. However, this embedding does \emph{not} compute the Unruh radiation correctly. A spiralling observer in Minkowski space observes non-thermal Unruh radiation \cite{Unruh:1998gq,Biermann:2020bjh}, unlike a geodesic observer in cylindrical spacetime.
} Thus, seemingly one can increase \(\alpha\) arbitrarily without penalty.

However, another quantum phenomenon threatens the ultrarelativistic observer: the Casimir effect. If we suppose for simplicity that matter living on the cylindrical time is conformal with central charge \(c\), then the Casimir effect produces a nonzero vacuum expectation value of the stress--energy density tensor \cite[§5.4.2]{DiFrancesco:1997nk}, which is as follows for an observer at rest:
\begin{equation}
    \begin{pmatrix}T_{\mu\nu}\end{pmatrix}=
    -\frac c{24\pi R^2}
    \begin{pmatrix}
        1&0\\0&1
    \end{pmatrix}.
\end{equation}
An ultrarelativistic observer moving at an angular velocity \(\omega<1/R\) therefore observes a Casimir stress--energy tensor
\begin{equation}
    \begin{pmatrix}T_{\mu\nu}\end{pmatrix}=
    -\frac c{24\pi R^2(1-\omega^2R^2)}
    \begin{pmatrix}
        1+\omega^2R^2&-2\omega R\\
        -2\omega R&1+\omega^2R^2
    \end{pmatrix},
\end{equation}
corresponding to a Casimir energy \(E_\mathrm{Casimir}\) and a Casimir momentum \(P_\mathrm{Casimir}\) given by
\begin{align}
    E_\mathrm{Casimir}&=-\frac{c(1+\omega^2R^2)}{12R(1-\omega^2R^2)},&
    P_\mathrm{Casimir}&=\frac{c\omega}{6(1-\omega^2R^2)}.
\end{align}
In particular, for \(\omega\) close to the upper bound \(1/R\), the rotating observer observes a Casimir momentum of
\begin{equation}
    P_\mathrm{Casimir} \approx \frac c{6R(1-\omega^2R^2)}.
\end{equation}
Thus, an observer who can only withstand momentum scales up to order \(E\gg1/R\) can only travel at angular momenta \(\omega\) such that
\begin{equation}
    1/(1-\omega^2R^2)\le ER/6c,
\end{equation}
and the time advantage is therefore bounded above by
\begin{equation}
    \alpha \lesssim \sqrt{ER/6c}.
\end{equation}
In particular, \(\ln\alpha\) does not have any terms of order \(\mathcal O(t)\), so this is less efficient than \eqref{eq:unruh-bound} over a long time period.

Of course, one can try to cancel the Casimir stress--energy by working in a spacetime filled with additional matter with positive energy fine-tuned to compensate for the Casimir effect. In this case, however, since spacetime is now filled with a medium, the observer and computer will now have a finite probability of being destroyed due to interaction with the medium; in particular, this does not allow for taking a \(\tau_\mathrm{obs}\to\infty\) limit to compare with \eqref{eq:unruh-bound}.

\subsection{Anti-de Sitter space}
Another example in which one can try to putatively violate \eqref{eq:unruh-bound} is the anti-de~Sitter space \(\operatorname{AdS}_d\), which is an example of a Malament--Hogarth spacetime \cite{1992FoPhL...5..173H,Earman_Norton_1993}. While the Unruh effect of anti-de~Sitter space can be correctly computed by the embedding method \cite{Deser:1997ri}, the embedding \(\operatorname{AdS}_d\hookrightarrow\mathbb R^{2,d-1}\) is into signature \((2,d-1)\) with \emph{two} time directions, rather than into \(\mathbb R^{1,d}\) as for de~Sitter space; this allows for the following construction, summarised in \cref{fig:AdS}.

Consider global coordinates in \(\operatorname{AdS}_d\) given by
\begin{equation}
    \mathrm ds^2 = -\left(1+(r/L)^2\right)\,\mathrm dt^2 + \frac{\mathrm dr^2}{1+(r/L)^2} + r^2\,\mathrm d\Omega_{d-2}^2,
\end{equation}
where the conformal boundary is located at \(r=\infty\).
In the embedding-space coordinates \((z^{-1},z^0,z^1,\dotsc,z^{d-1})=(z^{-1},z^0,\vec z)\), the space \(\operatorname{AdS}_d\) corresponds to the universal cover of the hyperboloid
\begin{equation}
    (z^{-1})^2+(z^0)^2-\|\vec z\|^2 = L^2,
\end{equation}
with the embedding map given by
\begin{align}
    z^{-1}&=\sqrt{L^2+r^2}\cos(t/L),&
    z^0&=\sqrt{L^2+r^2}\sin(t/L),&
    \vec z&=r\vec\Omega,
\end{align}
where \(\vec\Omega\) is the unit vector in \(\mathbb R^{d-1}\) given by the angular coordinates.

Suppose that the computer and the observer follow  timelike curves in the embedding space given by
\begin{align}
    \gamma_\mathrm{obs}(\tau_\mathrm{obs})&=\begin{pmatrix}\sqrt{L^2+a^{-2}}\cos\frac{\tau_\mathrm{obs}}{\sqrt{L^2+a^{-2}}}\\ 
    \sqrt{L^2+a^{-2}}\sin\frac{\tau_\mathrm{obs}}{\sqrt{L^2+a^{-2}}}\\a^{-1}\\
    0\\\vdots\\0\end{pmatrix},\\
    \gamma_\mathrm{comp}(\tau_\mathrm{comp})&=\begin{pmatrix}\sqrt{L^2+a^{-2}}\\a^{-1}\sinh(a\tau_\mathrm{comp})\\a^{-1}\cosh(a\tau_\mathrm{comp})\\0\\\vdots\\0\end{pmatrix}.
\end{align}
The computer and the observer both start from \((\sqrt{L^2+a^{-2}},0,a^{-1},0,\dotsc,0)\) at proper time \(\tau=0\).
The observer stays at constant radial coordinate and experiences zero temperature \cite{Deser:1997ri}, and
\begin{equation}
    t = \frac{\tau_\mathrm{obs}}{\sqrt{1+(aL)^{-2}}}.
\end{equation}
For the computer, the conversion from proper time \(\tau_\mathrm{comp}\) to global time \(t\) is
\begin{equation}
    t = L\arctan(z^0/z^{-1})
    =L\arctan\frac{a^{-1}\sinh(a\tau_\mathrm{comp})}{\sqrt{L^2+a^{-2}}},
\end{equation}
so that as \(\tau_\mathrm{comp}\to \infty\), the computer reaches the conformal boundary of anti-de~Sitter space at global time \(t=\frac12\pi L\). Once the computer reaches the boundary, it signals the observer with its results. Its lightlike signals reach the observer at \(r=1/a\) in finite coordinate time
\begin{equation}
    \Deltaup t = \int_{1/a}^\infty\frac{\mathrm dr}{1+(r/L)^2}
    =\frac12\pi L-L\arctan \frac{1}{La}.
\end{equation}
The computer experiences Unruh radiation of temperature \(T=a/2\pi\) (where \(a\) was defined above as the uniform acceleration of the trajectory in the embedding space), which remains bounded. Thus, if the computer can survive this temperature indefinitely, then the time advantage is infinite \cite{1992FoPhL...5..173H,Earman_Norton_1993}:
\begin{equation}
\begin{split}
    \alpha&=\lim_{\tau_\mathrm{comp}\to\infty}\frac{\tau_\mathrm{comp}}{\tau_\mathrm{obs}+\sqrt{1+(aL)^{-2}}\Deltaup t}\\
    &=\frac{\lim_{\tau_\mathrm{comp}\to\infty}\tau_\mathrm{comp}}{\left(\frac12\pi L+\left(\frac12\pi L-L\arctan (aL)^{-1}\right)\right)\sqrt{1+(aL)^{-2}}}
    =\infty.
    \end{split}
\end{equation}
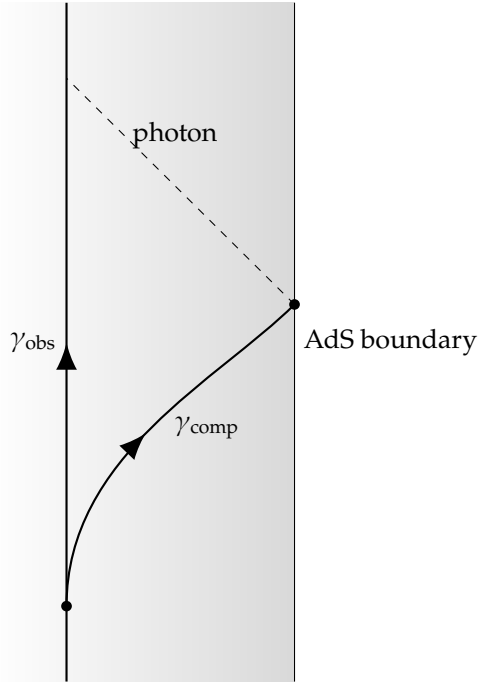
\begin{figure}
\begin{center}
\begin{tikzpicture}
  \coordinate (x) at (0,0);
  \coordinate (y) at ($(x)+(0,4)$);
  \coordinate (z) at ($(x)-(3,0)$);
  \coordinate (y') at ($(y)+(-3,3)$);
  \shade[shading=axis, right color=gray!30, left color=white] ($(x)-(0,1)$) rectangle (-4,8);
  \draw ($(x)-(0,1)$) -- ($(y)+(0,4)$) node [midway, right] {AdS boundary};

  \draw[thick, arrow]
    (z) .. controls ($(z)+(0,2)$) and ($(y)+(-1,-1)$) .. (y)
    node[midway, right=1ex] {\(\gamma_{\mathrm{comp}}\)};
  \draw [dashed] (y) -- (y') node [near end, right] {photon};
  \draw [thick, arrow] ($(z)-(0,1)$) -- ($(y')+(0,1)$)
      node[midway, left] {\(\gamma_{\mathrm{obs}}\)};

  \fill (z) circle (2pt);
  \fill (y) circle (2pt);

\end{tikzpicture}
\end{center}
\caption{
    Time-advantage scheme in anti-de Sitter space, which is Malament--Hogarth.
    The computer travels along a timelike curve \(\gamma_\mathrm{comp}\) with infinite proper time and experiences a finite temperature. The observer remains at a fixed radial coordinate along the timelike curve \(\gamma_\mathrm{obs}\). The computer has access to infinite proper time, but the observer only experiences a finite amount of proper time, seemingly leading to an infinite time advantage.}\label{fig:AdS}
\end{figure}

In this case, the culprit is the assumption that the computer can survive a finite temperature \(T>0\) indefinitely. This assumption is reasonable if the proper time \(\tau_\mathrm{comp}\) involved is finite, since it is exponentially unlikely that the computer will encounter a particle with energy \(E\gg T\) according to the Boltzmann distribution. On the other hand, if the proper time \(\tau_\mathrm{comp}\) involved is infinite, then at a finite temperature, for any energy \(E\), the computer will almost surely encounter a particle of energy \(E\).

How long, then, can the computer survive? If the computer can survive collisions with particles up to energy \(E\),
then an order-of-magnitude estimate given the Boltzmann distribution is
\begin{equation}
    \ln\tau_\mathrm{comp} \sim E/T = 2\pi E/a.
\end{equation}
Given this, the time advantage is
\begin{equation}
    \ln\alpha \sim 2\pi E/a.
\end{equation}
To compare with \eqref{eq:unruh-bound}, we divide by the proper time of the observer:
\begin{equation}
    \frac{\ln\alpha}{\tau_\mathrm{obs}} \sim \frac{2\pi E}{a(\pi L-\arctan (aL)^{-1})\sqrt{1+(aL)^{-2}}}
    \sim 4E - \frac{8EaL}\pi + \mathcal O((aL)^2)
    \le 4E.
\end{equation}
This is function is maximised at \(a=0\), with an upper bound \(4E\), so that we obtain qualitatively the same bound as \eqref{eq:unruh-bound}. (The coefficient \(4\) differs slightly from the coefficient \(\pi\) in \eqref{eq:unruh-bound}, but this is an apples-to-oranges comparison since  here we consider  energy rather  than temperature.)

Under the AdS/CFT correspondence, this process of a computer approaching (and burning up near) the horizon is dual to a `computer' that operates in a de~Sitter space boundary theory   \cite{Parikh:2012kg}. Using the de~Sitter slicing of anti-de Sitter space, the bulk computer corresponds to an  observer in the static patch of the boundary de Sitter geometry. The boundary computer must therefore accelerate to stay at a fixed radial coordinate in static coordinates for de~Sitter space and thereby experiences thermal Unruh radiation, which agrees with the temperature computed in the bulk, and therefore burns up in finite time.

\subsection{Eternal subextremal black holes}
Another example of a Malament--Hogarth spacetime is the maximal analytic extension of a subextremal Kerr--Newman black hole \cite{Etesi:2001ze}. For simplicity, we consider a four-dimensional subextremal Kerr black hole with zero charge (the presence of electric charge does not change the qualitative physics). The metric is given in Boyer--Lindquist coordinates as
\begin{multline}\label{eq:kerr}
    \mathrm ds^2 = 
    -(1-\frac{2mr}{\Sigma})\mathrm dt^2
    -\frac{2mra\sin^2\theta}\Sigma\,\mathrm dt\,\mathrm d\varphi
    +\frac\Sigma\Delta\,\mathrm dr^2+\Sigma\,\mathrm d\theta^2\\+
    \left(r^2+a^2+\frac{2mra^2\sin^2\theta}\Sigma\right)\sin^2\theta\,\mathrm d\phi^2,
\end{multline}
where
\begin{align}
    \Delta(r)&=r^2-2mr+a^2=\left(r-m-\sqrt{m^2-a^2}\right)\left(r-m+\sqrt{m^2-a^2}\right),\\
    \Sigma(r,\theta)&=r^2+a^2\cos^2\theta,
\end{align}
and the black-hole mass and angular momentum are \(m/G\) and \(am/G\), respectively; both \(m\) and \(a\) have units of length. The black hole is surrounded by an outer and an inner horizon, located at the roots of  \(\Delta(r)\), namely \(r_\pm=m\pm\sqrt{m^2-a^2}\) respectively; the singularity occurs at \(\Sigma(r,\theta)=0\). The outer horizon has temperature
\begin{equation}
    T_\mathrm{Kerr} = \frac{\sqrt{1-(a/m)^2}}{8\pi m\left(1+\sqrt{1-(a/m)^2}\right)}.
\end{equation}
As with all Malament--Hogarth spacetimes, a subextremal Kerr--Newman black hole is not globally hyperbolic; in this case, the failure of global hyperbolicity arises from the fact that the inner horizon is a Cauchy horizon, beyond which the Cauchy problem is not well-defined, similarly to anti-de~Sitter space.

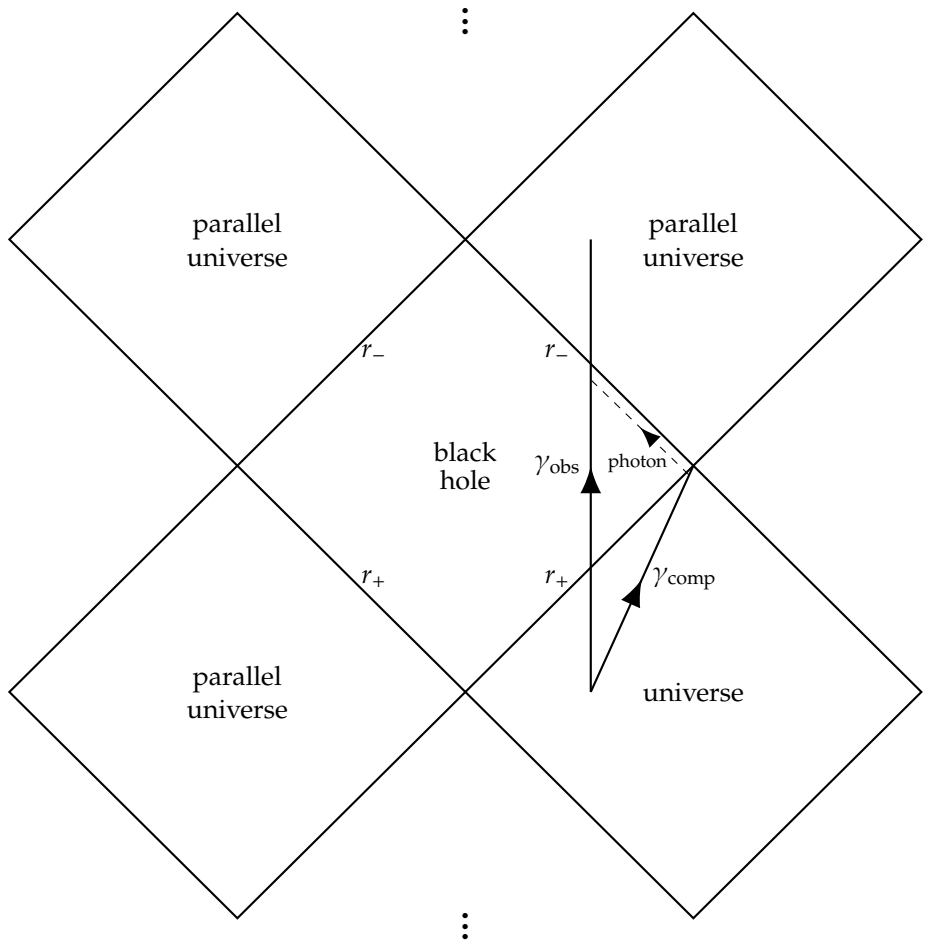
\begin{figure}
\centering
\begin{tikzpicture}[scale=3,
]

\draw[thick] (0,0) -- (1,1) -- (2,0) -- (1,-1) -- cycle;
\node at (1,0) {universe};
\draw[thick] (0,0) -- (-1,1) -- (0,2) -- (1,1);
\node at (0,1) {\shortstack{black\\hole}};
\node[left] at (0.5,0.5) {\(r_+\)};
\node[right] at (-0.5,0.5) {\(r_+\)};
\node[left] at (0.5,1.5) {\(r_-\)};
\node[right] at (-0.5,1.5) {\(r_-\)};
\draw[thick] (1,1) -- (2,2) -- (1,3) -- (0,2);
\node at (1,2) { \shortstack{parallel\\universe} };
\draw[thick] (0,0) -- (-1,-1) -- (-2,0) -- (-1,1);
\node at (-1,0) { \shortstack{parallel\\universe} };
\draw[thick] (-1,1) -- (-2,2) -- (-1,3) -- (0,2);
\node at (-1,2) { \shortstack{parallel\\universe} };

\node at (0,-1) { \LARGE\(\vdots\) };
\node at (0,3) { \LARGE\(\vdots\) };

\coordinate (x) at (0.55,0);
\draw[thick, arrow] (x) -- ($(x)+(0,2)$) node [midway, left] {\(\gamma_\mathrm{obs}\)};

\coordinate (y) at (1,1);
\draw[thick, arrow] (x) -- (y) node [midway, right] { \(\gamma_\mathrm{comp}\) };

\draw[dashed, arrow] ($(y)-(0.02,0.04)$) -- ($(x)+(0,1.5)-(0,0.12)$) node [very near start, left] {\scriptsize photon};

\end{tikzpicture}
\caption{The Penrose diagram of a Kerr black hole with outer horizon at \(r_+\) and inner Cauchy horizon at \(r_-\).
The observer, following a geodesic trajectory \(\gamma_\mathrm{obs}\), encounters the Cauchy horizon \(r_-\) in finite time, while the computer follows a timelike trajectory \(\gamma_\mathrm{comp}\) outside the black hole.
}\label{fig:kerr}
\end{figure}

In an eternal Kerr black hole, one has the following putative setup for an infinite time advantage \cite{1992FoPhL...5..173H,Earman_Norton_1993,Etesi:2001ze}, as illustrated in \cref{fig:kerr}. The observer dives into the black hole, crosses the outer horizon at \(r_+\), and reaches the Cauchy horizon at \(r_-\) in finite time, following the timelike curve \(\gamma_\mathrm{obs}\). In the meantime, the computer stays outside the black hole, following the timelike curve \(\gamma_\mathrm{comp}\), and sends signals into the black hole via photons. The proper time experienced by the computer \(\int_{\gamma_\mathrm{comp}}\mathrm d\tau\) is infinite; the observer can access the result of this infinite proper time's worth of computation in finite proper time when they cross the inner Cauchy horizon, so that the time advantage \(\alpha=\tau_\mathrm{comp}/\tau_\mathrm{obs}=\infty\) is infinite.
Furthermore, the timelike curves \(\gamma_\mathrm{obs}\) and \(\gamma_\mathrm{comp}\) can be arranged to be both geodesics \cite{Etesi:2001ze}, in which case neither the observer nor the computer encounters Hawking radiation from the black hole.

This analysis ignores black-hole evaporation; as \cite[§4]{Etesi:2001ze} notes, this construction assumes an \emph{eternal} black hole. A black hole could be eternal for kinematic reasons (e.g.\ conservation of a global or gauged charge) or for dynamical reasons (e.g.\ the black hole is immersed in a thermal bath in thermal equilibrium with it).
In the former scenario, for a black hole carrying a large amount of charge (either of a global or gauged symmetry), the endpoint of Hawking evaporation might be a remnant; if this remnant is macroscopic and admits an effective description in terms of general relativity, then the above scheme for infinite time advantage could go through.
However, in quantum gravity, several swampland conjectures (reviewed in \cite{Brennan:2017rbf,Palti:2019pca,Grana:2021zvf,Agmon:2022thq,Lehnert:2025izp}) exclude this possibility.
Specifically, the no-global-symmetry conjecture \cite{Banks:2010zn} states that there are no global charges;
the weak-gravity conjecture \cite{Arkani-Hamed:2006emk} implies that gauged charges can always be evaporated;
the cobordism conjecture \cite{McNamara:2019rup} even more generally implies that there are no kinematic obstacles to black hole evaporation since the interior of the black hole must be null-cobordant.

In light of the swampland conjectures, let us instead consider black holes made eternal by dynamical means. Concretely, one can keep the black hole eternal by continually feeding it, that is, putting the black hole in a thermal bath of the same temperature \(T_\mathrm{Kerr}\) as the black hole, so that thermal radiation falling into the black hole balances the Hawking radiation emitted from it.

In this revised scenario, at least the computer outside the outer event horizon is subject to interaction with a thermal bath of temperature \(T_\mathrm{Kerr}\) despite moving along a geodesic. Suppose that the computer can survive energy scales up to \(\Lambda_\mathrm{comp}\). Then, according to the Boltzmann distribution, the expected time scale \(\tau_\mathrm{comp}\) during which the computer survives intact is of order
\begin{equation}
    \ln\tau_\mathrm{comp}
    \sim \Lambda_\mathrm{comp}/T_\mathrm{Kerr} < \infty.
\end{equation}

To compute the time advantage, we must also compute \(\tau_\mathrm{obs}\), the proper time it takes for the observer to reach the Cauchy horizon. Geodesic motion in a Kerr spacetime is completely integrable \cite{Carter:1968rr} (reviewed in \cite[Ch.~4]{ONeill:2014rav}), thanks to the existence of a rank-two Killing tensor in addition to the timelike and azimuthal Killing vectors. For a timelike geodesic \(\gamma(\tau)\), parameterised by proper time, the quantities
\begin{align}
    E &= -g_{\mu0}\dot\gamma^\mu(\tau), & L &= g_{\mu3}\dot\gamma^\mu(\tau)
\end{align}
are constant, where \(g_{\mu0}\) and \(g_{\mu3}\) are picking out the components corresponding to the coordinates \(t\) and \(\varphi\), since \(\partial/\partial t\) and \(\partial/\partial\phi\) are Killing vectors of the Kerr spacetime \eqref{eq:kerr}; clearly \(E>0\), while \(L\) can have either sign. In addition, the geodesic \(\gamma(\tau)=\left(t(\tau),r(\tau),\theta(\tau),\varphi(\tau)\right)\) obeys the following differential equations (the radial equation and the colatitude equation, respectively) \cite[Th.~4.2.2]{ONeill:2014rav}:
\begin{align}
    \Sigma^2\dot r^2&=-\Delta(r^2+K)+\left((r^2+a^2)E-La\right)^2,\label{eq:kerr-radial}\\
    \Sigma^2\dot\theta^2&=K-a^2\cos^2\theta-\frac{\left(L-Ea\sin^2\theta\right)^2}{\sin^2\theta}.
\end{align}
In the above, \(K\ge0\) is a conserved quantity of the geodesic, the (shifted) \emph{Carter constant}, associated to a rank-two Killing tensor of the Kerr spacetime \eqref{eq:kerr}.

We wish to compute the proper time of observers that cross the Cauchy horizon \(r_-\), whose causal past includes the entirety of the exterior of the black hole; this corresponds to what are called `long flyby orbits of case B' in \cite[p.~247]{ONeill:2014rav}, in which case
\begin{align}
    0&<K,&1&\le E,&
    2mEr_-&<aL<2mEr_+,
\end{align}
as discussed in \cite{Etesi:2001ze}.

Suppose that the observer starts out near the outer horizon \(r=r_+\). Using the radial equation \eqref{eq:kerr-radial}, the time to trek between the outer and inner horizons is given by
\begin{equation}
    \tau_\mathrm{obs}=\int_{r_-}^{r_+}\dot r^{-1}\,\mathrm dr=
    \int_{r_-}^{r_+}\frac\Sigma{\sqrt{((r^2+a^2)E-aL)^2-\Delta(r^2+K)}}.\label{eq:kerr-observer-time-integral}
\end{equation}
For the interval \(r_-\le r\le r_+\), since \(\Delta=(r-r_-)(r-r_+)\le0\) in this interval, the integrand in \eqref{eq:kerr-observer-time-integral} is bounded above as
\begin{equation}
    \frac\Sigma{\sqrt{((r^2+a^2)E-aL)^2-\Delta(r^2+K)}}
    \le 
    \frac{r_+^2+a^2}{\sqrt{((r^2+a^2)E-aL)^2-\Delta(r_-^2+K)}},
\end{equation}
and
\begin{equation}
    \tau_\mathrm{obs}
    \le
    \frac{r_+^2+a^2}E\int_{r_-}^{r_+}\frac1{\sqrt{(r^2+a^2-aL/E)^2-\Delta(r_-^2+K)/E^2}}<\infty.
\end{equation}
In particular, by increasing \(E\) (with fixed \(L/E\) and \((r_-^2+K)/E^2\)), we can have \(\tau_\mathrm{obs}\propto E^{-1}\) be arbitrarily small --- intuitively, by going faster, one can trek from the outer horizon to the inner horizon arbitrarily quickly with respect to proper time.

The logarithmic time advantage per observer's proper time is therefore
\begin{equation}
    \frac{\ln\alpha}{\tau_\mathrm{obs}} \gtrsim \mathcal O(1)\frac{E\Lambda_\mathrm{comp}}{(r_+^2+a^2)T_\mathrm{Kerr}}
\end{equation}
(where the \(\mathcal O(1)\) factor depends on \(L/E\) and \((r_-^2+K)/E^2\)).
In particular, it can be made arbitrarily large. In order to compare to \eqref{eq:unruh-bound}, we would like to amortise in the limit \(\tau_\mathrm{obs}\to\infty\), so that the observer needs to repeat this trick of falling into black holes over and over again. This is possible if the observer can in fact successfully cross the Cauchy horizon, after which they enter a parallel universe where the trick can be repeated (\cref{fig:kerr}), in which case \eqref{eq:unruh-bound} would be violated.

The no-transmission principle of Engelhardt and Horowitz \cite{Engelhardt:2015gla} is a proposed quantum-gravity principle that directly rules out this scenario. According to this principle, (at least if one changes the scenario slightly to work with asymptotically anti-de~Sitter black holes) the observer's initial universe and the parallel universe that they enter correspond holographically to two different boundary conformal field theories, which should not be coupled to each other; an observer successfully crossing the Cauchy horizon would entail a signal being transmitted from one copy of the conformal field theory to the other, which is not allowed. In particular, the Cauchy horizon is (unlike the outer horizon\footnote{That is, at least if the firewall proposal \cite{Almheiri:2012rt} does not hold; if the firewall proposal does hold, the whole discussion of Kerr black holes is irrelevant.}) a true firewall, and an observer that tries to cross it is burnt up. This is corroborated by both numerical and rigorous results that the metric near the Cauchy horizon cannot be continuously twice differentiable under generic perturbations \cite{Brady:1995ni,Brady:1995un,Brady:1998ht,Dafermos:2017dbw} and that scalar modes' energies can diverge near the Cauchy horizon \cite{Dafermos:2003wr,Luk:2015qja} related to mass inflation \cite{Poisson:1989zz,Hamilton:2008zz}.
In other words, quantum gravity forbids observers from crossing the Cauchy horizon, so that the observer can use this trick only once (before meeting a swift fiery demise) and hence cannot violate \eqref{eq:unruh-bound} amortised over an arbitrarily long proper time.

Similarly, consider an extremal Kerr--Newmann black hole. In this case, the temperature \(T_\mathrm{Kerr}\) vanishes, so the computer need not suffer any defects from thermal radiation. (Depending on details of the theory, some extremal black holes may suffer from a superradiant instability, but this may possibly be cured if one fine-tunes the charges \cite{Furuhashi:2004jk,Hod:2012wmy,Huang:2015jza,Hod:2015hza,Mai:2021yny,Huang:2022nzm}; in string theory, one expects the extremal BPS black branes to be stable.) However, in this case the inner and outer horizons coincide, and the no-transmission principle then bars crossing the horizon.

\section{Limits on compressing memory into a finite region}\label{sec:space}
In addition to time, in order to perform universal computation, a computer needs access to space in the form of memory.
Following the previous section, which discusses limits on compressing runtime, here let us consider whether and how quantum-gravitational effects foil attempts to compress arbitrary amounts of memory in a finite spatial region.

At the level of purely classical non-gravitational physics, there is no upper bound to the information density that can be stored in a given spatial region \(\Sigma\); if the computer runs out of memory, it can simply construct new memory with a higher density within the bounded region \(\Sigma\). In quantum gravity, however, the situation is different. For spherical spatial regions, the number of nats\footnote{that is, the natural-logarithm analogue of bits} of information $I$ that can be stored in \(\Sigma\) is given by the Bekenstein--Hawking bound
\begin{equation}
    I \le \frac1{4M_\mathrm{Pl}^{d-2}}\operatorname{vol}(\partial\Sigma),
\end{equation}
or more generally its covariant refinement due to Bousso \cite{Bousso:1999xy} (reviewed in \cite{Bousso:2002ju}).
For this paper, it is more suggestive to look at the Bekenstein bound \cite{Bekenstein:1980jp}: if a spherical system in \(d\) spacetime dimensions has diameter \(\ell\) and energy \(\Lambda_\mathrm{comp}\), then the number \(N\) of possible states is bounded by
\begin{equation}\label{eq:bekenstein-bound}
    \frac{\ln N}\ell \le \pi\Lambda_\mathrm{comp}.
\end{equation}
This bound is saturated by Schwarzschild black holes in four dimensions.
This form is suggestively similar to \eqref{eq:unruh-bound}; in fact, since we have written \eqref{eq:bekenstein-bound} in terms of the diameter rather than the radius, even the \(\mathcal O(1)\) constant agrees.

In addition to black holes, however, given sufficiently exotic field content, one can try to engineer schemes for information storage beyond black holes and hence try to violate \eqref{eq:bekenstein-bound}. Below, we examine two such schemes and explain how the swampland species-scale  \cite{Palti:2019pca} and  distance conjectures \cite{Ooguri:2006in} frustrate such attempts.

\subsection{Storing memory using particle species}
One would be able to putatively circumvent the Bekenstein bound \eqref{eq:bekenstein-bound} by exploiting a large number of particle species.
Most straightforwardly, if one had \(N_\mathrm{species}\) weakly interacting particle species, such that the entropy of a gas containing only one species is \(S\), then a gas containing \(N_\mathrm{species}\) types of particles would have entropy \(S^{N_\mathrm{species}}\). For \(N_\mathrm{species}\) sufficiently large, this could exceed the Bekenstein bound \eqref{eq:bekenstein-bound}.
This observation is not new and is one of the motivations of the swampland species-scale conjecture \cite{Palti:2019pca}. According to it, if the computer operates in an effective field theory operating at energy scales up to \(\Lambda\), then the number of particle species \(N_\mathrm{species}\) in this effective field theory obeys the bound
\begin{equation}\label{eq:species-scale}
    N_\mathrm{species} \le (M_\mathrm{Pl}/\Lambda)^{d-2}.
\end{equation}
The above has been shown to be consistent with the Bekenstein bound in various physical situations. For instance, in \(d=4\), consider a computer of size of order \(D\) and at a temperature \(T\sim 1/D\), which is the Hawking temperature of a black hole of this size.
Suppose that the effective field theory in which the computer operates contains quantum chromodynamics with \(N_\mathrm c\) colours (and therefore \(N\sim N_\mathrm{c}^2\) species). Above the deconfinement temperature, this region can then encode up to
\begin{equation}
    S_\mathrm{QCD} \sim D^3T^3N_\mathrm{c}^2 \sim N
\end{equation}
nats of information \cite{Kaplan:2019soo}, whereas a black hole of the same size encodes
\begin{equation}
    S_\mathrm{BH} \sim (M_\mathrm{Pl}D)^2
\end{equation}
nats of information. Clearly, \(S_\mathrm{QCD}\) threatens to exceed \(S_\mathrm{BH}\) if the length scale \(D\) is small.
If the effective field theory is valid up to energy scale \(\Lambda\) (so that, in particular, \(E\lesssim\Lambda\)), then the smallest possible size of the black hole is \(D\gtrsim\Lambda^{-1}\). Thus, in order to obey the Bekenstein bound, one must have
\begin{equation}
    N_\mathrm{species}\lesssim (M_\mathrm{Pl}/\Lambda)^2
\end{equation}
in agreement with the species-scale conjecture \eqref{eq:species-scale} for \(d=4\) \cite{Kaplan:2019soo}.
In fact, the link between the species number \(N\) and entropy (and hence information and memory) has been made precise in  analogues of the laws of thermodynamics for the species number \cite{Cribiori:2023ffn}.

\subsection{Storing memory in vacuum moduli}
One might try to evade a species-based memory bound by encoding information in the value of a light modulus rather than in particle species.
Suppose that there exists a modulus, that is, a scalar field \(\phi\) with a flat potential, so that its expectation value \(\langle\phi\rangle\) can be set to an arbitrary value.  Assuming this is possible,  a computer that can either generate arbitrarily large $\Delta \phi = \phi_\mathrm{in}-\phi_\mathrm{out}$ \emph{or} resolve arbitrarily small field variations $\delta\phi$,  could in principle store an unbounded amount of information in the modulus field. This putatively would violate the Bekenstein bound \eqref{eq:bekenstein-bound}, similarly to the previous subsection.

\begin{figure}
\centering
\begin{tikzpicture}
\begin{axis}[
    hide axis,
    xmin = -2, xmax = 2,
    ymin = -0.1, ymax = 1.1,
    samples = 200,
    smooth,
    thick,
    width=10cm,
    height=6cm
]

\addplot[black, domain=-1.5:1.5] {exp(-1/(2.25-x^2))/(4*exp(-1/2.25))};
\addplot[black, domain=-2:-1.5] {0};
\addplot[black, domain=1.5:2] {0};

\end{axis}

\node at (4,0.5) { computer };
\node[above left] at (2.5, 1) { \shortstack{profile of\\modulus \(\phi\)} };
\node[above] at (4, 1.3) { \(\phi_\mathrm{in}\) };
\node[above] at (8, 0.4) { \(\phi_\mathrm{out}\) };

\draw[|<->|] (1,0) -- (2.5,0) node[midway, below=3pt] {\(w\)};

\end{tikzpicture}
\caption{
    To use the modulus \(\phi\) as a form of memory, the computer changes the expectation value \(\langle\phi\rangle\) near it with a small energy cost localised at the shell interpolating between \(\phi_\mathrm{in}\) and \(\phi_\mathrm{out}\).
    The approximate thickness of the shell is \(w\).
    The change of \(\langle\phi\rangle\) propagates outward at the speed of light.
}\label{fig:vev}
\end{figure}
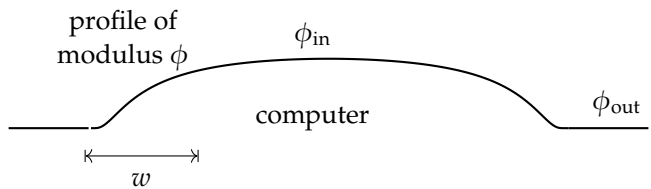

Let us first assume  \(\langle\Delta \phi\rangle\) is unbounded. Of course, with a spatially localised computer, one cannot change the expectation value of \(\phi\) everywhere simultaneously. At best, the computer could create a localised region where \(\phi\) takes a different value \(\phi_\mathrm{in}\) compared to the outside value \(\phi_\mathrm{out}\), as in \cref{fig:vev} (cf.\ \cite{Fischer:2024hqy}).
The energy density of the shell can be made small. If the thickness of the shell is \(w\), then the energy density of the shell is of order
\begin{equation}
    \rho \sim \|(\partial\phi)\|^2\, \sim (\phi_\mathrm{in}-\phi_\mathrm{out})^2w^{-2}.
\end{equation}
Thus, by taking $w$ sufficiently large, the energy density can be made arbitrarily small. However, avoiding gravitational collapse requires controlling the total energy of the configuration, not just the local density.  The energy incurred in creating a shell of width \(w\) and radius \(R\) is of the order \(\int\mathrm d^{d-1}x\,(\partial\phi)^2 \sim ((\phi_\mathrm{out}-\phi_\mathrm{in})/w)^2\cdot w\cdot R^2 = (\phi_\mathrm{out}-\phi_\mathrm{in})^2R^2/w\). To avoid gravitational collapse, we would like the Schwarzschild radius to be less than \(R\), which translates to \(M_\mathrm{Pl}^{2-d}(\phi_\mathrm{out}-\phi_\mathrm{in})^2R^2/w\lesssim R\) or, equivalently, \(|\phi_\mathrm{out}-\phi_\mathrm{in}|\lesssim M_\mathrm{Pl}^{d/2-1}\sqrt{w/R}\), and we must have \(w/R=\mathcal O(1)\) since \(w\) is the width of the shell and \(R\) is the radius of the shell.

Perhaps we have not been sufficiently ingenious and there is some more subtle mechanism by which the computer adiabatically  increments  $\phi_\mathrm{in}$, while avoiding gravitation collapse\footnote{To be clear, assuming a spatially bound region of influence this seems implausible in view of the classic gravitational collapse theorems \cite{Penrose:1964wq}.}. Even in this case, the swampland distance conjecture \cite{Ooguri:2006in} puts an upper bound on $\phi_\mathrm{in}$: as one moves a large distance in scalar field space, an infinite tower of states becomes exponentially light changing the physics of the computer. More precisely, for a proper field distance $\Delta\phi$, the mass scale of the tower behaves as $m \sim m_0 \exp(-\alpha \Delta\phi/M^{d/2-1}_{\mathrm{Pl}})$, with $\alpha = \mathcal{O}(1)$. As a result, after an order-one displacement in Planck units, the effective field theory breaks down due to the proliferation of light states.  A physical computer operates under  laws described by an effective field theory.  Upon creating a sufficiently large-value  memory state \(\langle\phi_\text{in}\rangle\),  the emergent  light tower of fields will generically couple to the computer and, when the effective field theory has broken down, one cannot guarantee that it  will function correctly. The \emph{refined} distance conjecture \cite{Blumenhagen:2018nts}, in \(d=4\), asserts that this breakdown occurs for $\Delta\phi \sim \mathcal{O}(M_{\mathrm{Pl}})$ so that the memory storage is again  bounded above by the Planck scale.

While the refined distance conjecture puts an upper bound to the range of \(\langle\phi\rangle\), it does not concern the precision with which the computer can usefully measure \(\langle\phi\rangle\). A notch on the metre stick carries much more information if one can measure its position in micrometres compared to if one only has centimetre precision.
Let the precision be \(\delta\phi\); then the vacuum expectation value of \(\phi\) carries \(\ln(M_\mathrm{Pl}/\delta\phi)\) nats of information.
If the computer has size of order \(D\) and operates at an energy of order \(E\), then the bound \eqref{eq:bekenstein-bound} together with the refined distance conjecture requires that
\begin{equation}
    \ln(M_\mathrm{Pl}/\delta\phi)\lesssim\mathcal O(1)DE,
\end{equation}
or, equivalently,
\begin{equation}\label{eq:precision}
    \delta\phi\gtrsim M_\mathrm{Pl}\exp(-\mathcal O(1)DE).
\end{equation}
We  are not aware of prior swampland discussion of bounds of this type in the literature. Perhaps predictably, \eqref{eq:precision} can also be interpreted heuristically as arising from gravitational backreaction. Resolving the modulus to accuracy \(\delta\phi\) requires extracting information through physical processes involving finite energy and time. A device of size \(D\) and energy \(E\) can perform at most \(\mathcal{O}(DE)\) independent operations without significant gravitational backreaction, and each operation yields at most \(\mathcal{O}(1)\) bits of information. Identifying this with \(S \sim \ln(M_{\mathrm{Pl}}/\delta\phi)\) again leads to \(\delta\phi \gtrsim M_{\mathrm{Pl}} \exp(-\mathcal{O}(1)DE)\), suggesting that the limitation on precision may be understood as a gravitational bound on information processing.

\newcommand\cyrillic[1]{\fontfamily{Domitian-TOsF}\selectfont \foreignlanguage{russian}{#1}}

\bibliographystyle{unsrturl}
\bibliography{biblio}
\end{document}